\documentclass[eqsecnum,aps,epsfig]{revtex4}
\usepackage{amsmath,amssymb,amsthm,bm}
\usepackage{psfrag}
\usepackage[dvips]{graphicx}
\usepackage[utf8]{inputenc} 
\usepackage{hyperref}

\begin{document}
\title{Influence of center vortex interactions on the static potentials}
\author{Seyed Mohsen Hosseini Nejad}
\email{smhosseininejad@semnan.ac.ir}
\affiliation{
Faculty of Physics, Semnan University, P.O. Box 35131-19111, Semnan, Iran}

\begin{abstract}

We analyze the static potentials induced by vacuum domains for various representations in SU($4$) Yang-Mills theory
within the framework of the domain structure model. By studying the interactions within the vacuum domains, we can uncover fundamental properties of the static potentials. It appears that attractions within the vacuum domains strongly adhere to Casimir scaling at intermediate distances. Conversely, the repulsions within the vacuum domains may decompose them into center vortices with the lowest magnitude of center vortex fluxes, thereby exhibiting $N$-ality at asymptotic distances.
\end{abstract}

\maketitle

\section{INTRODUCTION}\label{Sect0}

In non-perturbative QCD, explaining how color confinement arises is a challenging problem. Numerical simulations in SU($N$) lattice gauge theories, where there are non-trivial center elements, show that the static potentials between color sources are proportional to the eigenvalues of the quadratic Casimir operator at intermediate distances, known as the Casimir scaling hypothesis \cite{Deldar:1999vi,Bali:2000un,Piccioni:2005un}. At asymptotic distances, the string tension of color charges matches that of the lowest representation with the same $N$-ality \cite{Kratochvila:2003zj}, where $N$-ality classifies the representations of a gauge group \cite{Kratochvila:2003zj}. Additionally, the static potentials must be everywhere convex and without concavity \cite{Bachas:1985xs}. Numerical simulations \cite{DelDebbio:1996lih,Langfeld:1997jx,DelDebbio:1997ke,Langfeld:1998cz,Engelhardt:1999fd,Kovacs:1998xm} have indicated that center vortices \cite{tHooft:1977nqb,Vinciarelli:1978kp,Yoneya:1978dt,Cornwall:1979hz,Mack:1978rq,Nielsen:1979xu,Dehghan:2024rly,Asmaee:2021xkm,Golubich:2021kjc,Schweigler:2012ae,Hollwieser:2012kb} which are quantized magnetic flux tubes could account for the quark confinement. The thick center vortex model \cite{Faber:1997rp} disscused for SU($2$) can expalin the basic properties of the static potentials and the model can be generalized to $N \ge 2$ \cite{Deldar:2001,Deldar:2007}. Furthermore, for G$_2$ gauge theory there is the identity element only. However, lattice simulations in G$_2$ show a linear potentials for all representations at intermediate distances which satisfy the Casimir scaling \cite{Holland:2003jy,Greensite:2006sm,Olejnik:2009jr,Pepe:2007}. In Refs. \cite{Greensite:2006sm,Liptak:2008gx,Nejad:2014tja}, a domain structure is assumed in the vacuum for G$_2$ and SU($N$) gauge theories. The QCD vacuum is filled with domains of center vortex type corresponding to 
the non-trivial center elements and of vacuum type corresponding to the identity center element. The domain structure model of the confinement mechanism could be able to explain both Casimir scaling at intermediate distances and $N$-ality dependence at large distances. The model \cite{Faber:1997rp,Greensite:2006sm,Engelhardt:1999wr,Engelhardt:2003wm,Deldar:2010hw,Deldar:2011fh,Deldar:2009aw,Nejad:2014hka} indicates that the area law for Wilson loops is due to the linking of the loops with the thick center domains. We used one type of vacuum domain in $G_2$ \cite{Deldar:2011fh,Nejad:2014hka} and another one for SU($2$) and SU($3$) \cite{Nejad:2014tja,Nejad:2019}.

In this paper, motivated by the lattice simulations in G$_2$, which indicate the identity element plays an important role in confinement, we study both types of vacuum domains in SU($4$) gauge group. Analyzing only vacuum domains, it seems that one can get to basic properties of the static potentials, Casimir scaling at the intermediate distances and $N$-ality at large distances. We investigate the static potentials induced by two types of vacuum domains individually. Interactions within the vacuum domains may show the stable configurations. It seems that attractions between center vortices in the vacuum domains lead to almost the exact Casimir scaling in the static potentials at the intermediate regime while repulsions between center vortices in vacuum domains exhibit $N$-ality at the asymptotic regime.

In section \ref{Sect1}, we briefly explain the domain structure model. As introduced in \cite{Greensite:2006sm}, the color magnetic fields of each domain are located in a square regime where the total magnetic flux over each domain corresponds to an element
of the gauge-group center. Within the framework of the domain model, in section \ref{Sect2}, we analyze the static potentials in the lowest representations and their ratios induced by two types of vacuum domains in SU($4$) gauge theory. The attractions and repulsions in the vacuum domains are studied in subsections \ref{subsect1} and \ref{subsect2}. The main points of our study are summarized in section \ref{Sect3}. 

\section{The model of Casimir scaling and N-ality}\label{Sect1}

The domain structure model can explain some features of the confining force obsereved in lattice simulations \cite{Faber:1997rp,Greensite:2006sm}. In the model, the QCD vacuum is filled with domains. In SU($N$), the domains consist of $N-1$ types of center vortices corresponding to 
the non-trivial center elements $z_n=\exp(i2\pi n/N)$ where $n=1,...,N-1$, and vacuum type corresponding to the trivial center element $z_n=1$ ($n=0,N$). 
The effect of a domain on a planar Wilson loop from the representation $r$ is to multiply the loop by a group factor as

\begin{equation}
W_r(C)\longrightarrow {G}_r(\alpha^{(n)})\; \mathbf{I}_{d_r} W_r(C),
\label{W}
\end{equation}
where
\begin{equation}
{G}_r(\alpha^{(n)})=\frac{1}{d_r}Tr\left(\exp\left[i\vec{\alpha}^{(n)}\cdot\vec{\mathcal{H}}\right]\right),
\end{equation}
$d_r$ is the dimension of the representation $r$, $\{\mathcal{H}_i\}$ are generators of the Cartan subalgebra, the angle $\alpha^{(n)}_C(x)$ depends on both the Wilson loop $C$ and the position $x$ of the domain, and $\mathbf{I}_{d_r}$ is the $d_r\times d_r$ unit matrix. If a domain is entirely contained within the loop area, then
\begin{equation}
\exp\left[i\vec{\alpha}^{(n)}\cdot\vec{\mathcal{H}}\right]=(z_n)^{k_{r}}\mathbf{I}_{d_r},
\label{max_alpha}
\end{equation}
where $k_{r}$ is the $N$-ality of representation $r$. Using this constraint, one can obtain the maximum value of the angle ${\alpha}^{(n)}_{max}$. Furthermore, if the domain is outside the loop, it has no effect on the loop. The quark potential induced by the domains is as follows \cite{Faber:1997rp,Greensite:2006sm}:
\begin{equation}
\label{potential}
V_r(R) = -\sum_{x}\ln( 1 - \sum_{n} f_{n}
[1 - {\mathrm {Re}}{G}_{r} (\vec{\alpha}^{(n)}_{C}(x))]),
\end{equation}
where $f_n$ is the probability that any given plaquette is pierced by an nth domain. The square ansatz for the angle $\alpha^{(n)}_C(x)$ was introduced by Greensite $\it{et~ al.}$ \cite{Greensite:2006sm}. Each
domain, with cross section $A_d$, is divided to subregions of area $l^2\ll A_d$ which $l$ is a short correlation length. The color magnetic fluxes in subregions $l^2$ fluctuate randomly and 
almost independently, making the color magnetic fluxes in neighboring subregions $l^2$ uncorrelated. The only constraint is that the total color magnetic fluxes of the subregions 
must correspond to an element of the gauge group center. The square ansatz for the angle is as follows:  
\begin{equation}
\label{Sansax}
         \vec{\alpha}^{(n)}_C(x)\cdot\vec{\alpha}^{(n)}_C(x) = \frac{A_d}{ 2\mu} \left[
\frac{A}{ A_d} - \frac{A^2}{ A_d^2} \right]
                         + \left(\alpha^{(n)}_{max} \frac{A}{ A_d}\right)^2,
\end{equation}
where $A$ is the cross section of the center domain overlapping with the minimal area of the Wilson loop and $\mu$ is a free parameter. The cross section 
of a domain is an $L_d \times L_d$ square. Now one should take two intervals for the square ansatz: 

\begin{equation}\label{eq:phi-pl0}
\vec{\alpha}^{(n)}_C(x)\cdot\vec{\alpha}^{(n)}_C(x)=\begin{cases}\beta_1(x)&0 \le R\le L_d,\\ 
                \beta_2(x)&L_d\le R,\end{cases}
\end{equation}

where

\begin{equation}\label{eq:beta_1}
\beta_1(x)=\begin{cases}\frac{{L_d}^2}{ 2\mu} \left[
\frac{y{L_d}}{ {L_d}^2} - \frac{({y{L_d})}^2}{ {L_d}^4} \right]
                         + \left(\alpha^{(n)}_{max} \frac{{y{L_d}}}{ {L_d}^2}\right)^2&-\frac{{L_d}}{2} \le x\le -\frac{{L_d}}{2}+R,\\ 
                \frac{{L_d}^2}{ 2\mu} \left[
\frac{RL_d}{ {L_d}^2} - \frac{{RL_d}^2}{ {L_d}^4} \right]
                         + \left(\alpha^{(n)}_{max} \frac{{RL_d}}{ {L_d}^2}\right)^2&-\frac{{L_d}}{2}+R \le x\le \frac{{L_d}}{2},\\ 
                \frac{{L_d}^2}{ 2\mu} \left[
\frac{y{L_d}}{ {L_d}^2} - \frac{({y{L_d})}^2}{ {L_d}^4} \right]
                         + \left(\alpha^{(n)}_{max} \frac{{y{L_d}}}{ {L_d}^2}\right)^2&\frac{{L_d}}{2} \le x\le R+\frac{{L_d}}{2},\end{cases}
\end{equation}

\begin{equation}\label{eq:beta_2}
\beta_2(x)=\begin{cases}\frac{{L_d}^2}{ 2\mu} \left[
\frac{y{L_d}}{ {L_d}^2} - \frac{({y{L_d})}^2}{ {L_d}^4} \right]
                         + \left(\alpha^{(n)}_{max} \frac{{y{L_d}}}{ {L_d}^2}\right)^2&-\frac{{L_d}}{2} \le x\le \frac{{L_d}}{2},\\ 
                (\alpha^{(n)}_{max})^2&\frac{{L_d}}{2} \le x\le R-\frac{{L_d}}{2},\\ 
                \frac{{L_d}^2}{ 2\mu} \left[
\frac{y{L_d}}{ {L_d}^2} - \frac{({y{L_d})}^2}{ {L_d}^4} \right]
                         + \left(\alpha^{(n)}_{max} \frac{{y{L_d}}}{ {L_d}^2}\right)^2&R-\frac{{L_d}}{2} \le x\le R+\frac{{L_d}}{2},\end{cases}
\end{equation}

and

\begin{equation}\label{eq:beta_1}
y(x)=\begin{cases}R-x+\frac{{L_d}}{2}&|R-x| \le |x|,\\  
                x+\frac{{L_d}}{2}&|R-x| > |x|.\end{cases}
\end{equation}

The range of the position $x$ of the domain $\it{i.e.}$ $-\frac{{L_d}}{2} \le x\le R+\frac{{L_d}}{2}$ is restricted over all plaquettes that the domains overlap to the minimal area of the Wilson loop.

\begin{figure}[h!]
\centering
a)\includegraphics[width=0.25\columnwidth]{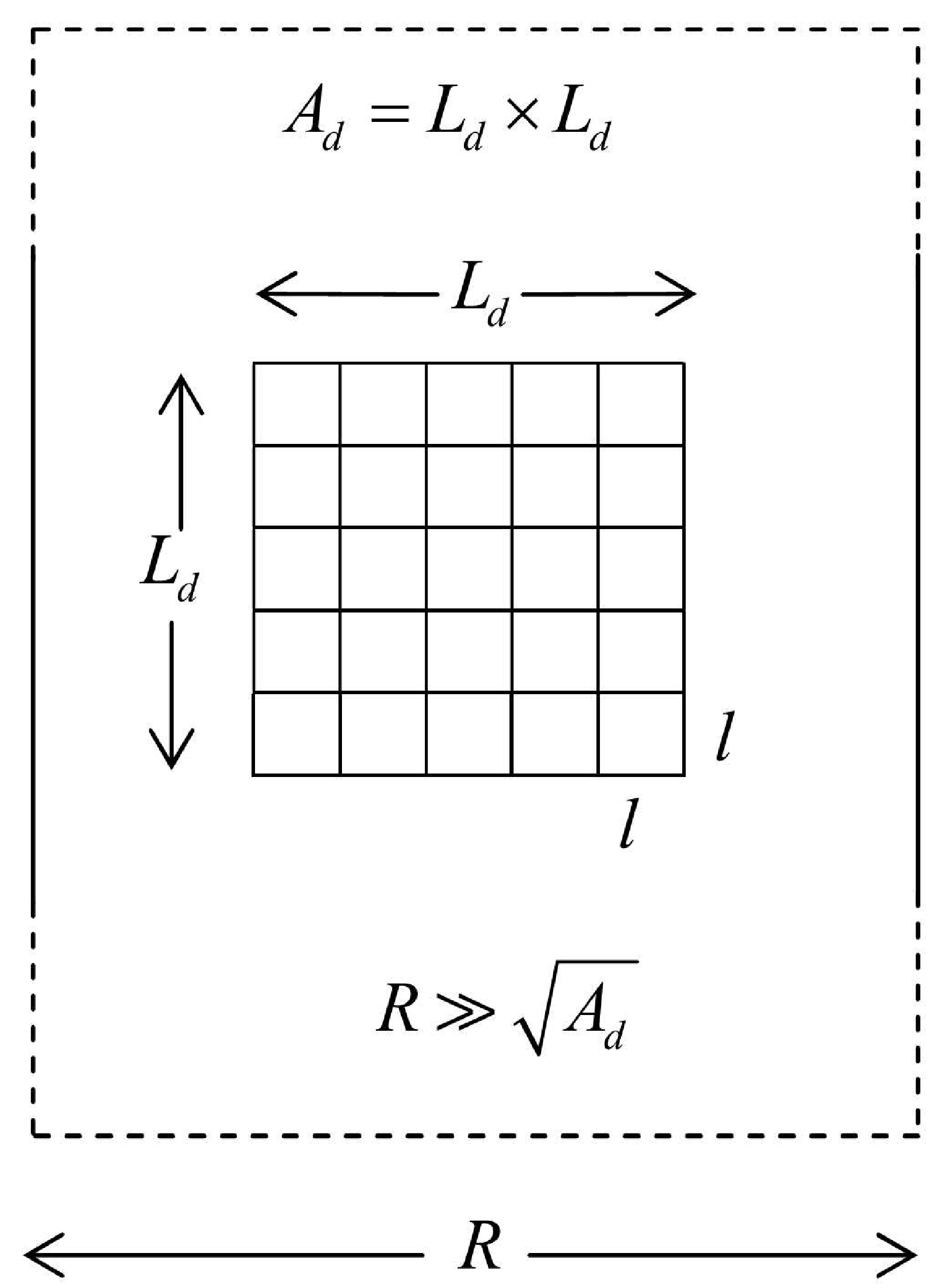}
\hspace{5em}
b)\includegraphics[width=0.135\columnwidth]{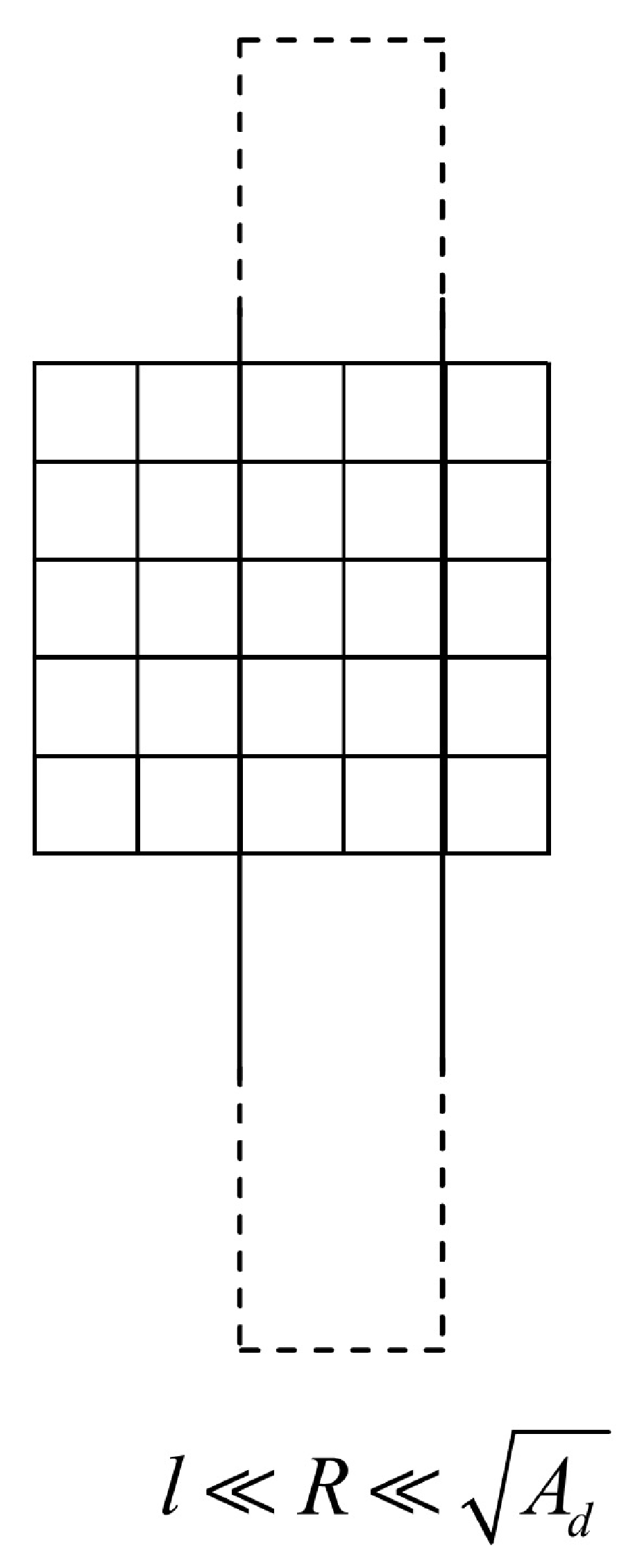}
\caption{The cross section of the center domain overlaps with the minimal area of the Wilson loop a) for large loops where almost all domains are entirely enclosed
by the loop and the static potentials are governed by the $N$-ality b) for medium size loops where $L_d \times R$ of the cross section of almost all domains overlap with the loops and the potentials are qualitatively in agreement with Casimir scaling.  }\label{fig:1}
\end{figure}

Figure \ref{fig:1} shows schematically the interaction between the angle of square ansatz and the Wilson loop for two intervals. In particular, consider large loops, $R\gg\sqrt{A_d}$, which the linear potential depends on the $N$-ality $k_{r}$ of the representation, almost every vortex which affects the loop is entirely enclosed
by the loop, and for these domains $\alpha^{(n)}_C(x) \approx \alpha^{(n)}_{max}$, as shown in Fig. \ref{fig:1}a). Only those domains near the perimeter have
$\alpha^{(n)}_C(x)$ different from $\alpha^{(n)}_{max}$, this is a negligible fraction of
the total contribution. For medium size loops, where $\alpha^{(n)}_C(x)$ is also typically small, the static potential $V_r(R)$ of the representation $r$ will be linearly rising for distances $l\ll R\ll\sqrt{A_d}$ and approximately proportional to its quadratic Casimir. In this regime, the cross section $L_d \times R$
of a domain locates within the loop, as shown in Fig. \ref{fig:1}b). The contribution of those domains near the perimeter overlapping with the minimal area less than $L_d \times R$ of their cross sections is small. 

In the next section, we focus on the contribution of the static potentials between color sources induced by vacuum domains for SU($4$)
gauge theory.

\section{Static potentials induced by SU($4$) vacuum domains}\label{Sect2}

\begin{figure}[h!]
\centering
\includegraphics[width=0.42\columnwidth]{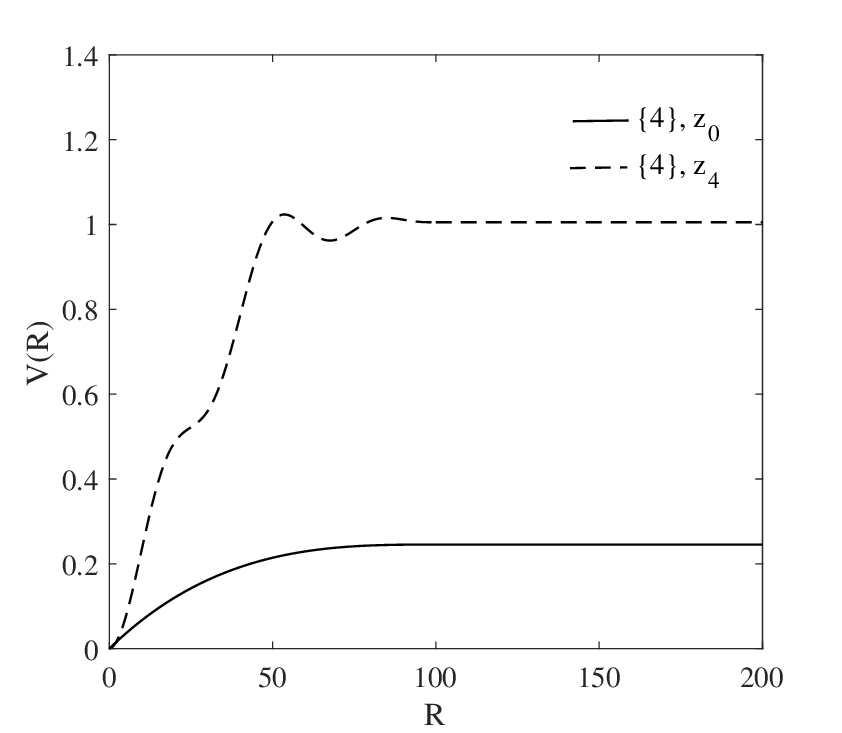}
\caption{ The static potentials induced by two types of vacuum domains individually for the fundamental representation $\{4\}$. The concavity is observed for the static potential induced by $z_4$ vacuum domains while there is no this artifact in the potential obtained by $z_0$ vacuum domains. The free parameters are $L_{d}=100$, $f_{0}=0.03$, $f_{4}=0.01$, and $L^{2}_{d}/(2\mu)=4$.}\label{fig:2}
\end{figure}

According to the homotopy group
\begin{equation}
\Pi_1[SU(4)/\mathbb{Z}(4)] = \mathbb{Z}(4),
\end{equation}
this implies that the SU($4$) gauge theory has center domains corresponding to the center elements in $\mathbb{Z}(4)$ as:

\begin{equation}\label{center_elements}
         z_0=e^0,~~~
 \\
         z_1=e^{i\pi/2},~~~
 \\
         z_2=e^{i\pi},~~~
 \\ 
         z_3=e^{i3\pi/2},~~~
 \\
         z_4=e^{i2\pi},
\end{equation}

where there are three types of center vortices corresponding to 
the non-trivial center elements of $z_n$ ($n=1,2,3$) and two types of vacuum domains corresponding to the trivial center element $z_n=1$ ($n=0,4$). In some literatures, vortex of type $z_1$ and type $z_3$ have phase
factors which could be considered complex conjugates of one another ($z_3=z_1^*$) and therefore two vortices may be observed as the
same type of vortex but with magnetic flux pointing in opposite directions. In addition, in the same method, two types of vacuum domains may be regarded as the same type.  Without this constraint, vortex fluxes of three types of center vortices and two types of vacuum domains are different and we analyze the behavior of these center domains on static potentials. Using Eq. (\ref {potential}), the static potential induced by each type of center domain is as follows:
\begin{equation}\label{potential_square_ansatz}
V_r(R) =- \sum^{{ {L_d}/2 + R}}_{{x=-{L_d}/2}} \ln[(1-f_n) + f_n{\mathrm {Re}}{G}_r(\vec{\alpha}^{(n)}_{C}(x))]. 
\end{equation}
 Now, we analyze the static potentials induced by the vacuum domains. Using Eq. (\ref {max_alpha}), one can obtain the maximum value of the angle ${\alpha}^{(n)}_{max}$. The square ansatz given in Eq. (\ref {Sansax}) for the angles corresponding to two types of vacuum domains for all representations is:
\begin{equation}
         (\alpha^{(0)}_C(x))^{2} = \frac{A_d}{ 2\mu} \left[
\frac{A}{ A_d} - \frac{A^2}{ A_d^2} \right],~~~
 \\
         (\alpha^{(4)}_C(x))^{2} = \frac{A_d}{ 2\mu} \left[
\frac{A}{ A_d} - \frac{A^2}{ A_d^2} \right]
                         + \left({4\pi}{\sqrt{6}}\frac{A}{ A_d}\right)^2.                          
\end{equation}
The free parameters $L_{d}$, $f_{0}$, $f_{4}$, and $L^{2}_{d}/(2\mu)$ 
 are taken to be $100$, $0.03$, $0.01$, and $4$, respectively. The correlation length is chosen $l=1$ and therefore the static potentials are linear from the beginning ($R=l$). Figure \ref{fig:2} depicts the potentials induced by two types of vacuum domains individually for
the representation $\{4\}$ (fundamental). 
The concavity is appeared in the static potential induced by vacuum domains corresponding to
 $z_4$ while there is no this artifact in the potential obtained by vacuum domains of type $z_0$. The group factor ${G}_r(\alpha^{(n)})$ of the static potentials provides detailed information about the center domains. The real part of the group factors for the lowest representations of SU($4$), used in the static potential given in Eq. (\ref{potential_square_ansatz}), can be calculated as follows:

\begin{equation}
    {\mathrm {Re}}{G}_{\{4\}}(\alpha^{n})=\frac{1}{4}[3cos(\frac{\alpha^{n}}{2\sqrt{6}})+cos(\frac{3\alpha^{n}}{2\sqrt{6}})],   
\label{group-4}                      
\end{equation}
\begin{equation}
    {\mathrm {Re}}{G}_{\{6\}}(\alpha^{n})=cos(\frac{\alpha^{n}}{\sqrt{6}}),  
\label{group-6}                       
\end{equation}
\begin{equation}
    {\mathrm {Re}}{G}_{\{15\}}(\alpha^{n})=\frac{1}{15}[9+6cos(\frac{2\alpha^{n}}{\sqrt{6}})].  
\label{group-15}                       
\end{equation}
 
\begin{figure}[h!]
\centering
a)\includegraphics[width=0.42\columnwidth]{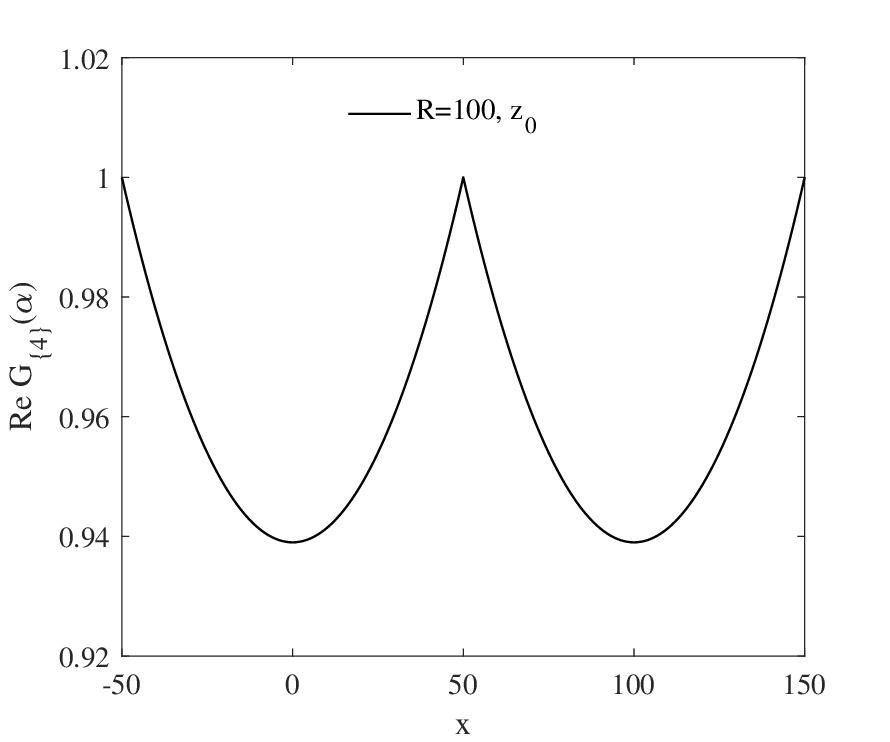}
b)\includegraphics[width=0.42\columnwidth]{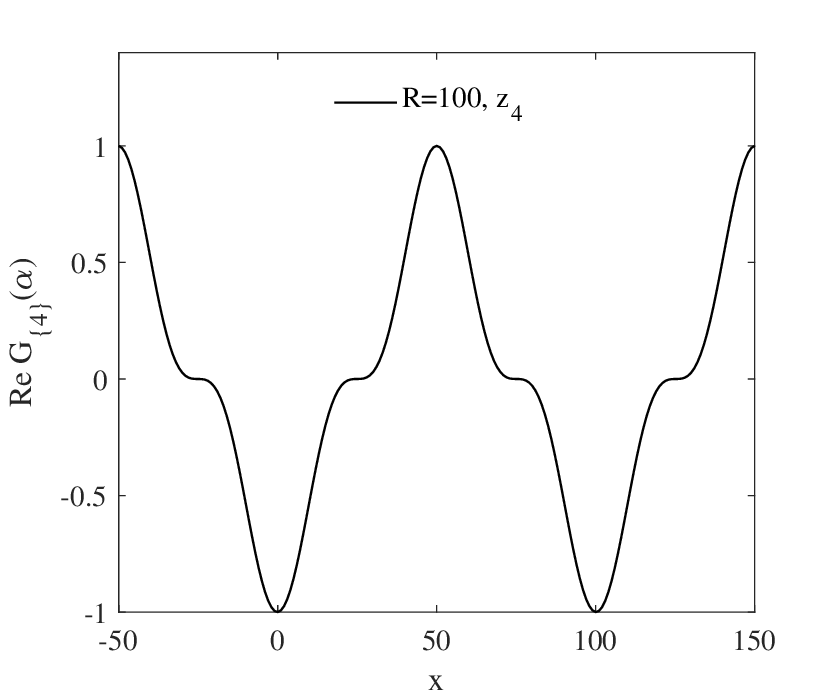}

\caption{The group factor ${\mathrm {Re}}G_{\{4\}}(\alpha)$ versus $x$ for the large distance with $R=100$ a) for the $z_0$ vacuum domains. b) for the $z_4$ vacuum domains. The time-like legs of the Wilson loop are located at $x = 0$ and $x = 100$. The group factor of the $z_0$ vacuum domains changes smoothly close to trivial value $1$ around any time-like leg and the one of $z_4$ vacuum domain includes the values $0$ and $-1$ around any time-like leg. The free parameters are $L_{d}=100$ and $L^{2}_{d}/(2\mu)=4$. }\label{fig:3}
\end{figure}

\begin{figure}[h!]
\centering
\includegraphics[width=0.42\columnwidth]{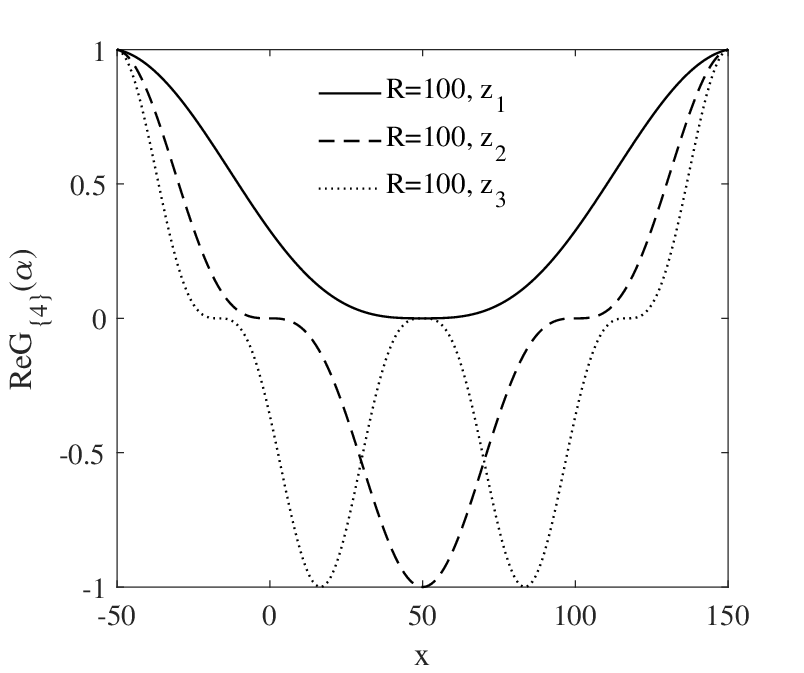}
\caption{The group factor ${\mathrm {Re}}G_{\{4\}}(\alpha)$ versus $x$ for the large distance with $R=100$ for the $z_1$, $z_2$, and $z_3$ center vortices. The time-like legs of the Wilson loop are located at $x = 0$ and $x = 100$. The group factor of $z_1$ ($z_3$) center vortices when the vortex core is located entirely within the Wilson loop is equal to $0$ and the one of $z_2$ center vortices when all of the vortex is pierced with the loop is equal to $-1$. As shown in Fig. \ref{fig:3}, these two values, $0$ and $-1$, appear in the group factor induced by $z_4$ vacuum domain while the group factor of the $z_0$ vacuum domains changes smoothly close to trivial value $1$ around any time-like leg. It seems that the $z_4$ vacuum domain is decomposed into the $z_1$, $z_2$, and $z_3$ center vortices. Additionally, it seems that $z_0$ vacuum domain, like $z_4$ domain, is constructed of the center vortices but with the opposite orientations. The free parameters are $L_{d}=100$ and $L^{2}_{d}/(2\mu)=4$.}\label{fig:4}
\end{figure}

Figure \ref{fig:3} depicts the group factors corresponding to two types of vacuum domains for the fundamental representation at large distances ($R=100$).  
The group factor of $z_0$ vacuum domain changes smoothly close to trivial value $1$ around any time-like leg and the one of $z_4$ vacuum domain is included the values $0$ and $-1$ around any time-like leg. 
To understand these group factors, we plot those of SU($4$) center vortices in Fig. \ref{fig:4}. As shown, the group factors for first and third types of center vortices, $z_1$ and $z_3$, when the vortex cores are located entirely within the Wilson loop are equal to $0$ and the one for $z_2$ vortex type, entirely enclosed
by the Wilson loop, reaches $-1$. It seems that $z_4$ vacuum domain of SU($4$) is constructed of the center vortices with the same flux orientations. Therefore, $z_4$ vacuum domain can be characterized by the center element $z_1^4$, $z_2^2$, or $z_1z_3$.
In addition, $z_0$ vacuum domain, which its group factor changes smoothly close to trivial value $1$, like $z_4$ one, is constructed of the center vortices but with the opposite orientations. As a result, $z_0$ vacuum domain can be characterized for example by the center element $z_1z_1^*$. 

Therefore, there are interactions between center vortices within two types of the vacuum domains.

 The QCD vacuum could be described in terms of a
Landau-Ginzburg model of a dual superconductor which is similar to type II superconductors but the roles of the electric and magnetic
fields, and electric and magnetic charges, have been interchanged \cite{Hooft,Mandelstam}. In the type II superconductors, the interaction between vortices as the magnetic flux lines is repulsive while the one between vortex and antivortex with opposite directions of the magnetic flux lines is attractive \cite{Kramer:1971zza,Chaves}.

Therefore, it seems that vortices in the $z_4$ vacuum domain repel each other while there is attractions between the vortex and anti-vortex in $z_0$ vacuum domain. 

According to the numerical lattice calculations for the G$_2$ gauge group which has only one trivial center element, the static potentials in different representations 
grow linearly at intermediate distances where the slopes are in the agreement with the Casimir scaling \cite{Holland:2003jy,Greensite:2006sm,Olejnik:2009jr,Pepe:2007}. Therefore, it is interesting to understand the role of the vacuum domains corresponding to the trivial center element in the static potentials.

In the next two subsections, we investigate the role of two types of vacuum domains in the Casimir Scaling and $N$-ality properties of the static potentials for SU($4$) gauge theory.

\subsection{Attractions in $z_0$ vacuum domains and static potentials }\label{subsect1}

\begin{figure}[h!]
\centering
a)\includegraphics[width=0.42\columnwidth]{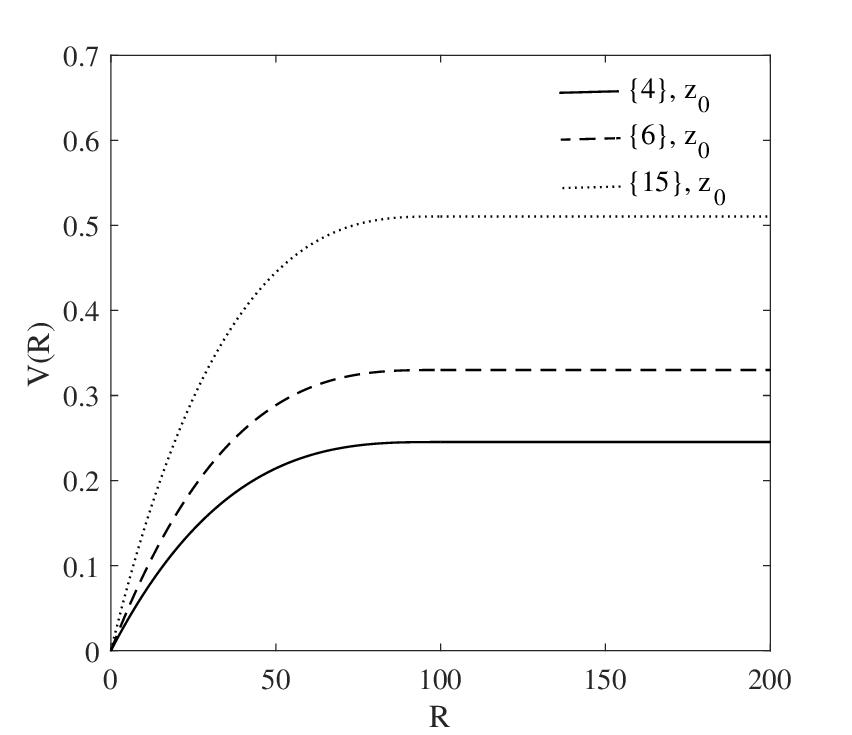}
b)\includegraphics[width=0.42\columnwidth]{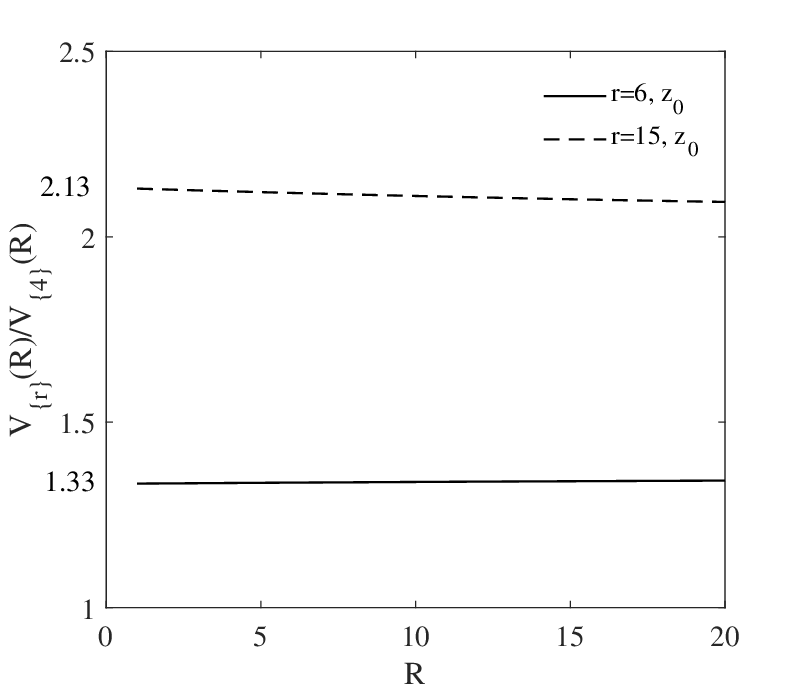}
\caption{a) The static potentials induced by $z_0$ vacuum domains. The potentials are linear in the range $R\in [0,20]$ where the $z_0$ vacuum domain is partially located
inside the Wilson loop. At large distances, the potentials are screened
at large distances where the vacuum domain is located
completely inside the Wilson loop. b) Potential ratios 
$V_{\{r\}}(R)/V_{\{4\}}(R)$ induced by $z_0$ vacuum domains at the intermediate distances. These potentials agree with almost exact Casimir scaling where center flux located in the medium size loops is typically small. The free parameters are chosen to be $L_{d}=100$, $f_{0}=0.03$, and $L^{2}_{d}/(2\mu)=4$.}\label{fig:5}
\end{figure}

In $z_0$ vacuum domains, two oppositely oriented center vortices attract each other, causing their cores to overlap and some of the magnetic flux in each vortex to be annihilated. Consequently, the magnetic flux of $z_0$ vacuum domain linked to different medium size loops is approximately zero and their group factors are close to one. The static potentials $V_{r}(R)$ induced by $z_0$ vacuum domains, for the $\{4\}$ (fundamental), $\{6\}$ and $\{15\}$ (adjoint) representations for the range $R\in [0,200]$ 
are plotted in Fig. \ref{fig:5} a). 
At intermediate distances, the potentials are linear in the range $R\in [0,20]$. Figure \ref{fig:5} b) plots the potential ratios $V_{\{6\}}(R)/V_{\{4\}}(R)$ and $V_{\{15\}}(R)/V_{\{4\}}(R)$ for $z_0$ vacuum domains. The potential ratios start out at the Casimir ratios: 
\begin{equation}
\frac{C_{\{6\}}}{C_{\{4\}}}=1.33, ~~~~~~~~~~~~~~~~~~~~~~~\frac{C_{\{15\}}}{C_{\{4\}}}=2.13.
\label{casimir}
\end{equation}

For the range $R \in [0,20]$, the potential ratios $V_{\{6\}}(R)/V_{\{4\}}(R)$ and  $V_{\{15\}}(R)/V_{\{4\}}(R)$ induced by $z_0$ vacuum domains change very slowly and agree with almost exact Casimir scaling.

\begin{figure}[h!]
\centering
\includegraphics[width=0.28\columnwidth]{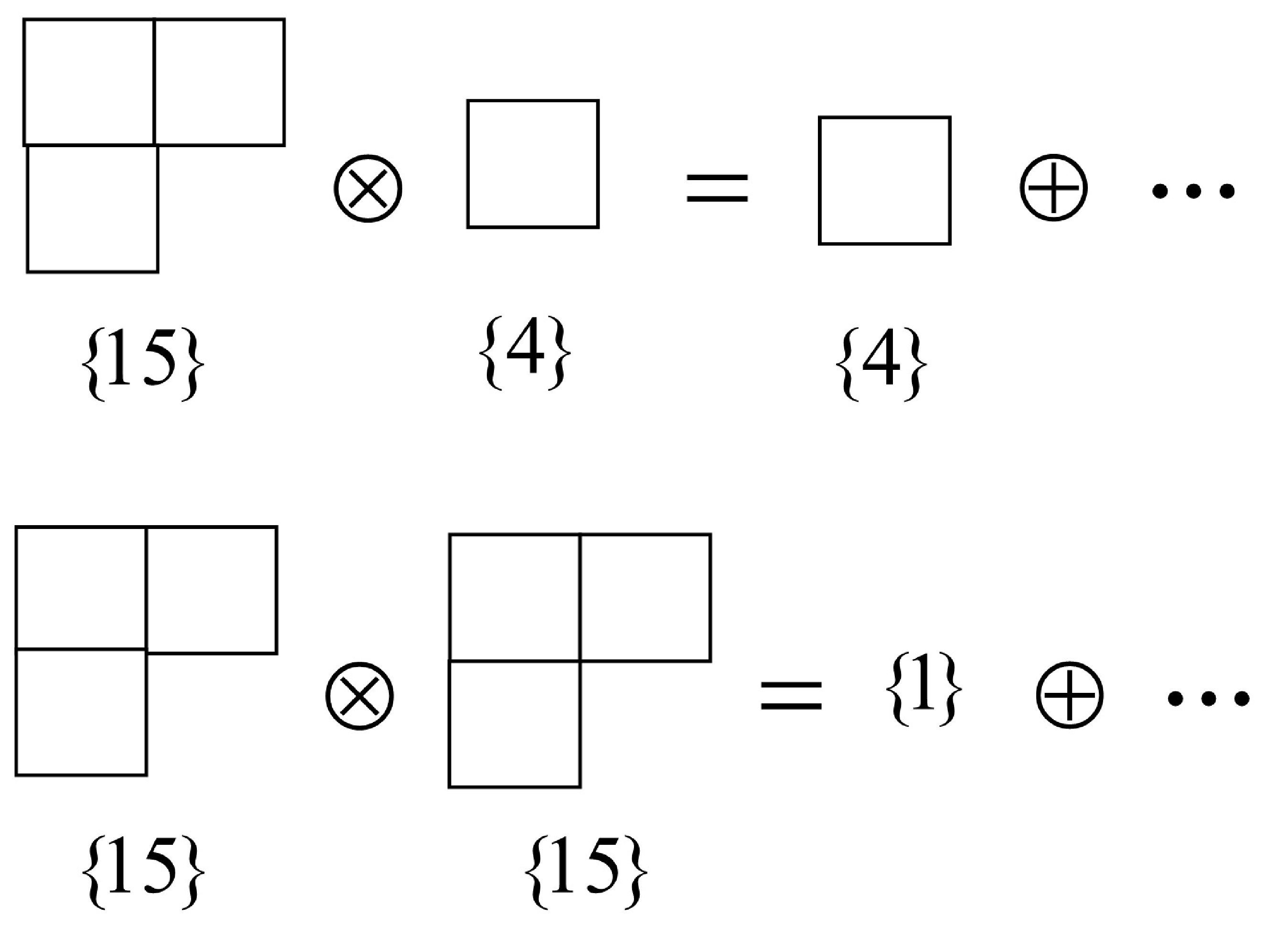}
\caption{The tensor product of the Young diagrams of the $\{4\}$, $\{15\}$ representations with a gluon $\{15\}$. Combining the fundamental representation $\{4\}$ with a gluon, we get again to the representation $\{4\}$. An adjoint charge $\{15\}$ combining with a gluon can form a color-singlet $\{1\}$. The representations $\{4\}$ and $\{15\}$ are the lowest-dimensional representation with the $N$-ality equal 1 and 0, respectively. The slops of the static potentials for the large distances should agree with the tensor product of a representation with a gluon, showing $N$-ality dependence.}\label{fig:6}
\end{figure}

\begin{figure}[h!]
\centering
\includegraphics[width=0.52\columnwidth]{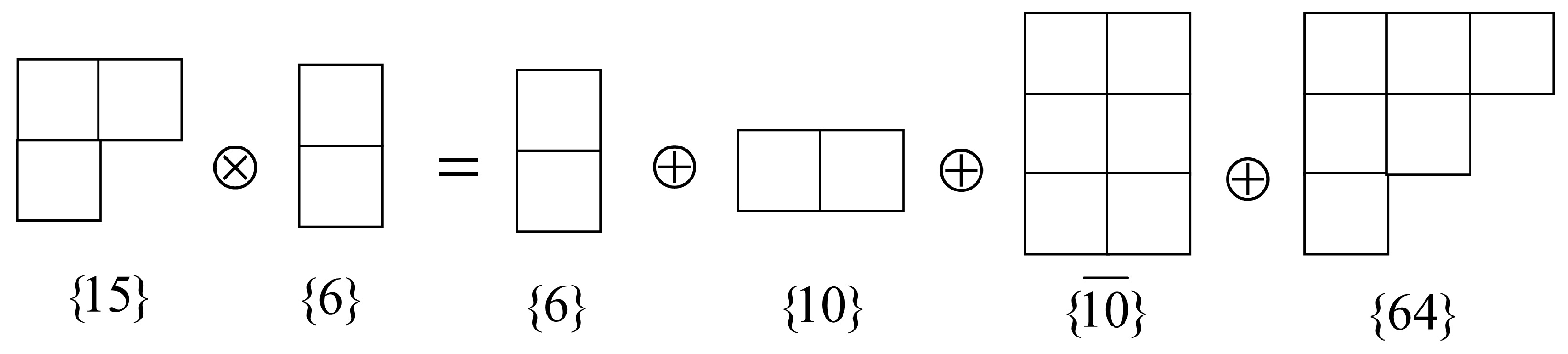}
\caption{The tensor product of the Young diagram of the representation $\{6\}$ with a gluon, $\{15\}$. Combining the representation $\{6\}$ with a gluon, is not possible to obtain $\{4\}$ or $\{1\}$ representation and we get again to the representation $\{6\}$. This indicates that the representation $\{6\}$ is also the lowest-dimensional representation with the $N$-ality equal to 2.}\label{fig:7}
\end{figure}

As shown in Fig. \ref{fig:5} a), at large distances, $R\geq 100$, the color screening is appeared for the static potentials induced by $z_0$ vacuum domains for all representations while $N$-ality should classify the representations of the gauge group. At large distances, when the energy between two static sources is equal to or greater than twice the 
gluon mass, a pair of gluons is created in the vacuum. These gluons combine with the initial sources, reducing the string tensions of the representations to the lowest-dimensional representation with the same $N$-ality. The tensor products of the Young diagrams of the analyzed representations for the potentials with a gluon are shown in Figs. \ref{fig:6} and \ref{fig:7}. 

\begin{figure}[h!]
\centering
a)\includegraphics[width=0.10\columnwidth]{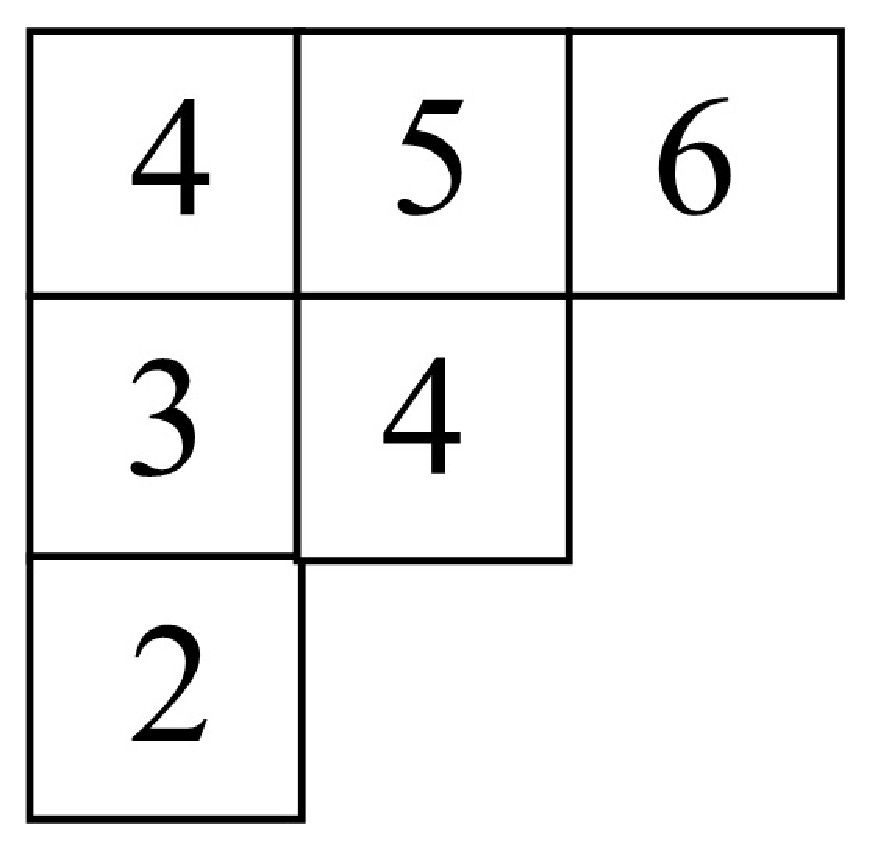}
\hspace{5em}
b)\includegraphics[width=0.10\columnwidth]{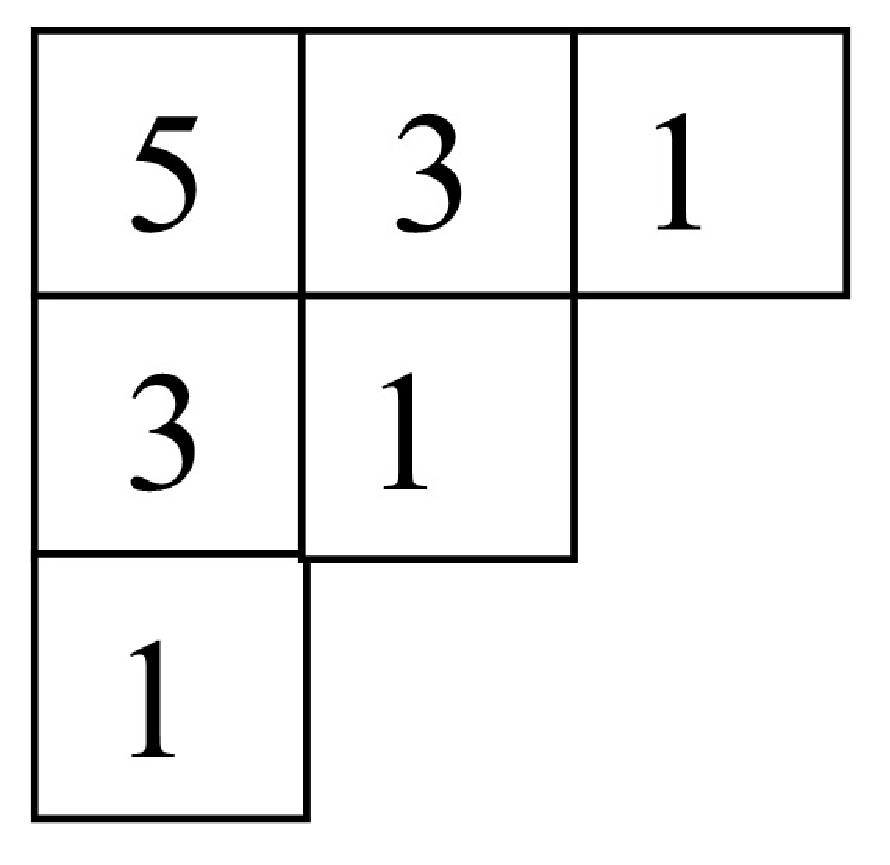}
\caption{The dimension of a Young diagram is determined by the following procedure. As an example, we obtain the 64-dimensional representation of an SU($4$) diagram obtained in the tensor product of Fig. \ref{fig:7}. a) the numbers of the boxes are given for this SU($4$) diagram where for an SU($N$) diagram, insert $N$, $N+1$, $N+2$, ... in the boxes along the first row and decrease the numbers by step of 1 along the vertical columns. For the SU($4$) Young diagram, the numerator $n$, as the product of all these numbers, is $n=6 \times 5 \times 4 \times 4 \times 3 \times 2= 2880$. b) The hook numbers of the SU($4$) diagram given in a). Draw a line from the right, entering the row in which the box is found.  On entering the box the line turns downward and passes down the column until it exits the diagram. The value of the hook for this box is the total number of boxes the line passes through. For the Young diagram, the denominator $d$, as the product of all the hooks, is $d=1 \times 3 \times 5 \times 1 \times 3 \times 1= 45$. The dimension of the representation is given by the ratio of the numerator $n$ and denominator $d$, $Dim=n/d=2880/45=64$.}\label{fig:8}
\end{figure} 

As shown in Fig. \ref{fig:6}, combining the fundamental representation $\{4\}$ with a gluon, one get again to the representation $\{4\}$. Therefore the representation $\{4\}$ is the lowest-dimensional representation with the $N$-ality equal 1. Additionally, an adjoint representation $\{15\}$ combining with a gluon can form a color-singlet $\{1\}$ and therefore the static potential for the adjoint representation with zero $N$-ality should be screened at large distances. In Fig. \ref{fig:7}, we illustrate the tensor product of the Young diagrams of the $\{6\}$ representation with a gluon $\{15\}$ where it is not possible to obtain $\{4\}$ or $\{1\}$ and we get again to the representation $\{6\}$. This indicates that the representation $\{6\}$ is also the lowest-dimensional representation with the $N$-ality equal to 2. Consequently, we should observe three slops of the potentials at large regime. In Fig. \ref{fig:8}, the dimensions of the Young diagrams are determined. Using the SU($4$) Young diagram with 64-dimensional representation, we show the procedure.

Thus, by considering the attractions between center vortices, the static potentials strongly agree with Casimir scaling at intermediate regimes, while they destroy $N$-ality at large distances.

In the next subsection, we analyze the role of $z_4$ vacuum domains in the $N$-ality property of the static potentials. 

\subsection{Repulsions in $z_4$ vacuum domains and static potentials }\label{subsect2}

 As argued, it seems that $z_4$ vacuum domain is constructed of center vortices with the same flux orientations, causing them to repel each other. Such $z_4$ vacuum domains are characterized by the center elements $z_4=z_1^4$, $z_4=z_2^2$, or $z_4=z_1z_3$. One may conclude that $z_4$ vacuum domain does not form a stable configuration and each center vortex within the $z_4$ vacuum domain should be considered as a single vortex in the model. As a result, the Yang–Mills vacuum in stead of the $z_4$ vacuum domain is filled with $z_1$, $z_2$, and $z_3$ vortices. 

\begin{figure}[h!]
\centering
\includegraphics[width=0.42\columnwidth]{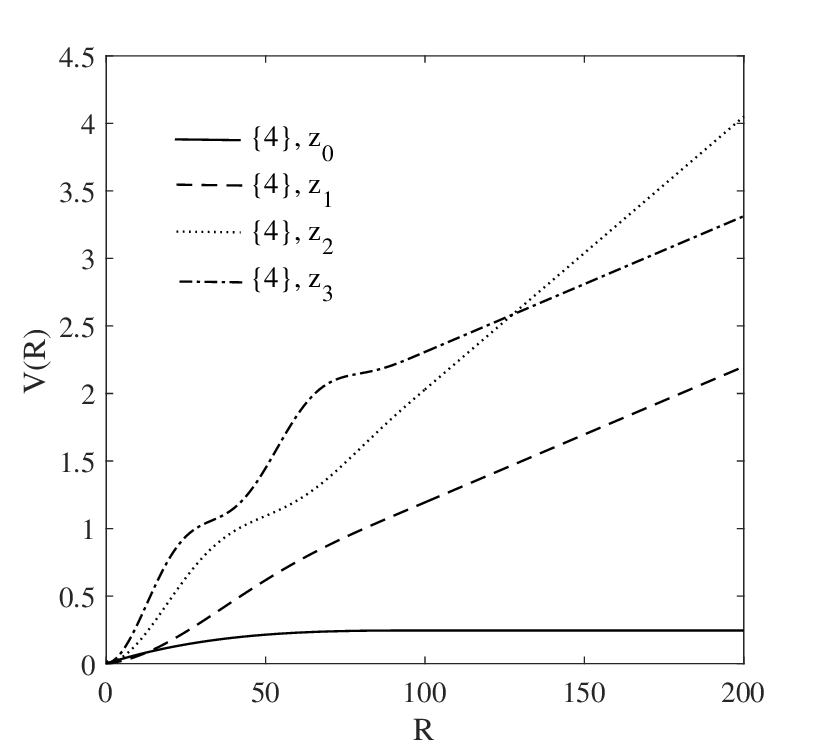}
\caption{ The static potentials induced by SU($4$) center domains individually for the fundamental representation. Concavity is observed in the potentials induced by $z_2$ or $z_3$ center vortices while this artifact does not appear in the potentials obtained from $z_1$ center vortices or $z_0$ vacuum domains. The free parameters are $L_{d}=100$, $f_{0}=0.03$, $f_{1}=f_{2}=f_{3}=0.01$, and $L^{2}_{d}/(2\mu)=4$.}\label{fig:9}
\end{figure}

Now, we analyze the static potentials induced by each of these center vortices. In Fig. \ref{fig:9}, we show the static potentials corresponding to SU($4$) center domains individually for the fundamental representation for the range $R\in [0,200]$.
\begin{figure}[h!]
\centering
a)\includegraphics[width=0.42\columnwidth]{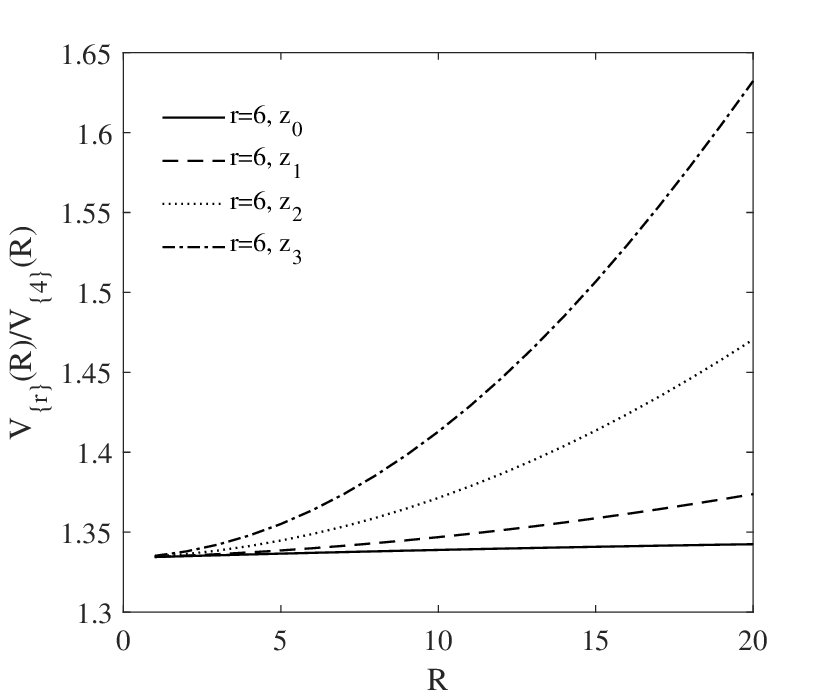}
b)\includegraphics[width=0.42\columnwidth]{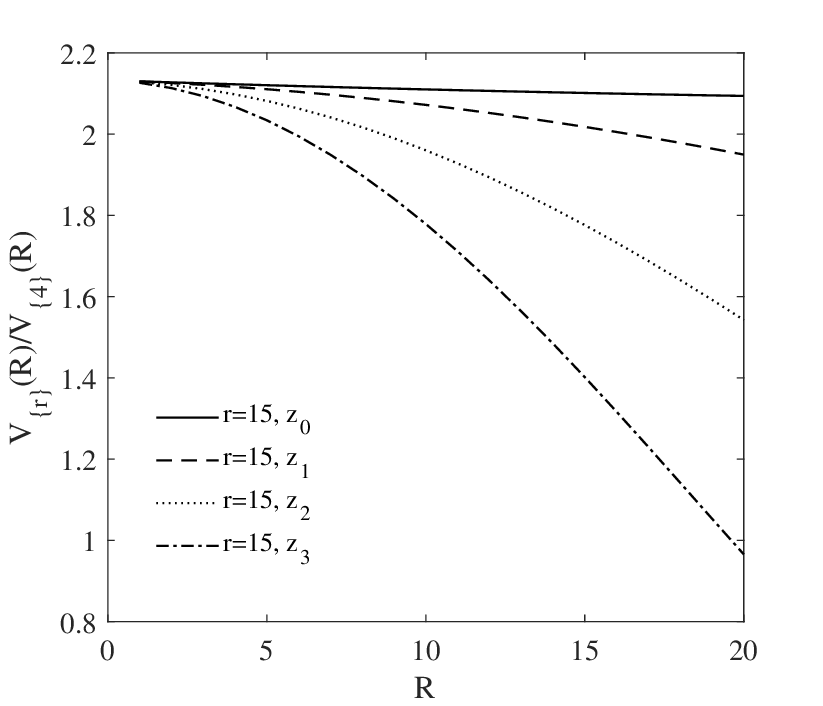}
\caption{The potential ratios of ${V_{\{r\}}(R)}/{V_{\{4\}}(R)}$ induced by SU($4$) center domains individually for the ${\{6\}}$ and ${\{15\}}$ representations. The potential ratios of the $z_0$ vacuum domains remain almost constant compared to the Casimir ratios. The potential ratios of the $z_1$ center vortices change very slowly compared to the Casimir ratios while the deviations from the exact Casimir scaling are much greater for those induced by $z_2$ or $z_3$ center vortices. The free parameters are $L_{d}=100$, $f_{0}=0.03$, $f_{1}=f_{2}=f_{3}=0.01$, and $L^{2}_{d}/(2\mu)=4$.}\label{fig:10}
\end{figure}
Again, concavity is observed for the potential induced by $z_2$ or $z_3$ center vortices while this artifact is absent in the static potentials obtained from $z_1$ center vortices or $z_0$ vacuum domains. Additionally, Fig. \ref{fig:10} plots the potential ratios $V_{\{6\}}(R)/V_{\{4\}}(R)$ and $V_{\{15\}}(R)/V_{\{4\}}(R)$ for each of SU($4$) cener vortices compared with those of $z_0$ vacuum domains for the range $R \in [0,20]$. The potential ratios induced by $z_0$ vacuum domains remain almost constant compared to the Casimir ratios. The potential ratios induced by $z_1$ center vortices change very slowly compared with the Casimir ratios while the deviations from the exact Casimir scaling are much greater for those induced by $z_2$ or $z_3$ center vortices. 
\begin{figure}[h!]
\centering
a)\includegraphics[width=0.42\columnwidth]{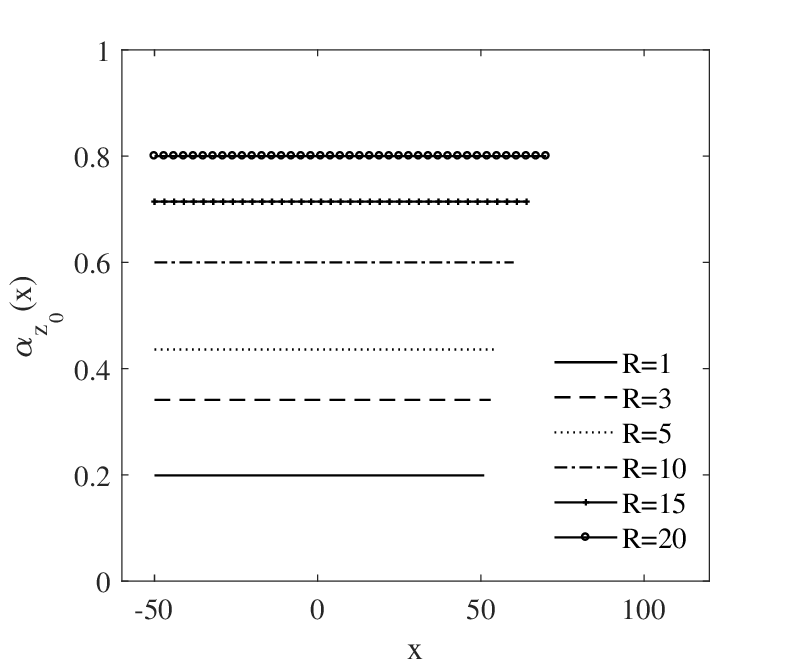}
b)\includegraphics[width=0.42\columnwidth]{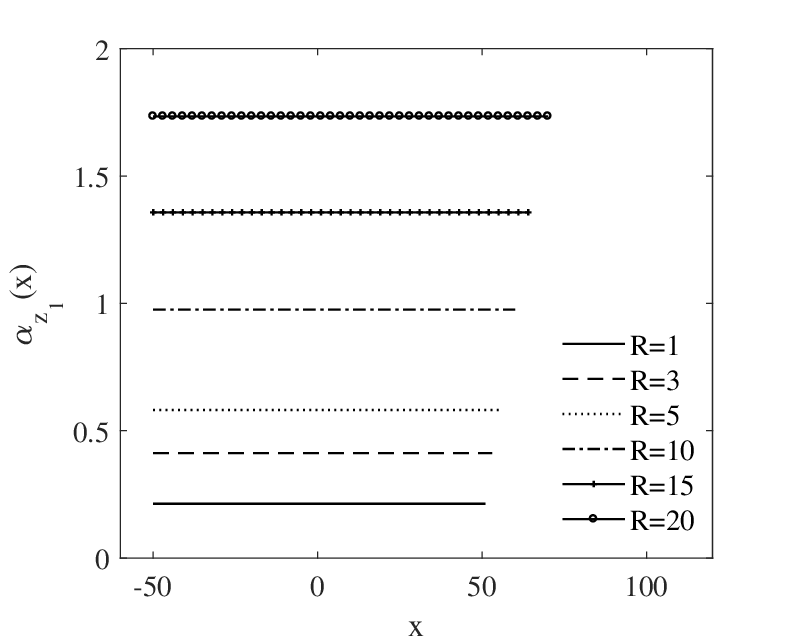}
c)\includegraphics[width=0.42\columnwidth]{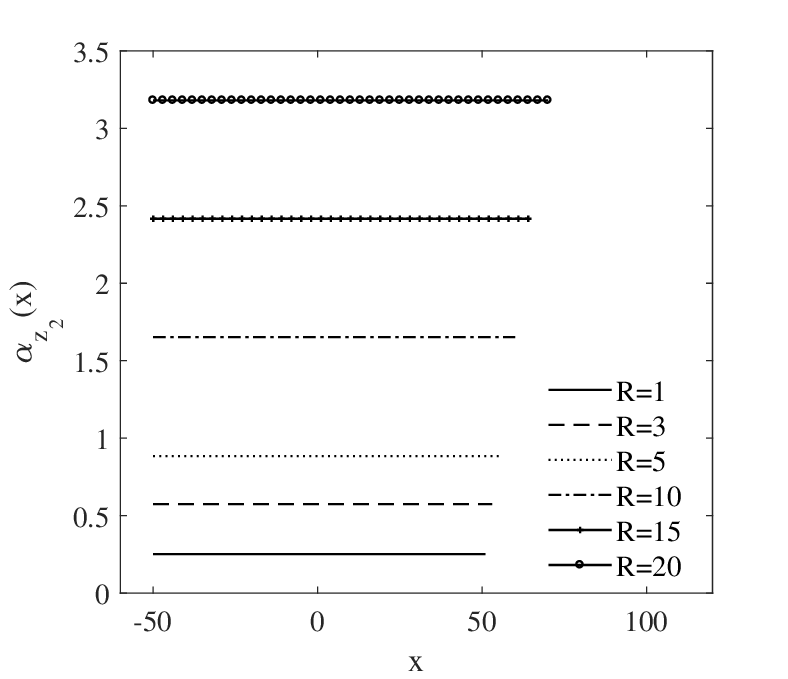}
d)\includegraphics[width=0.42\columnwidth]{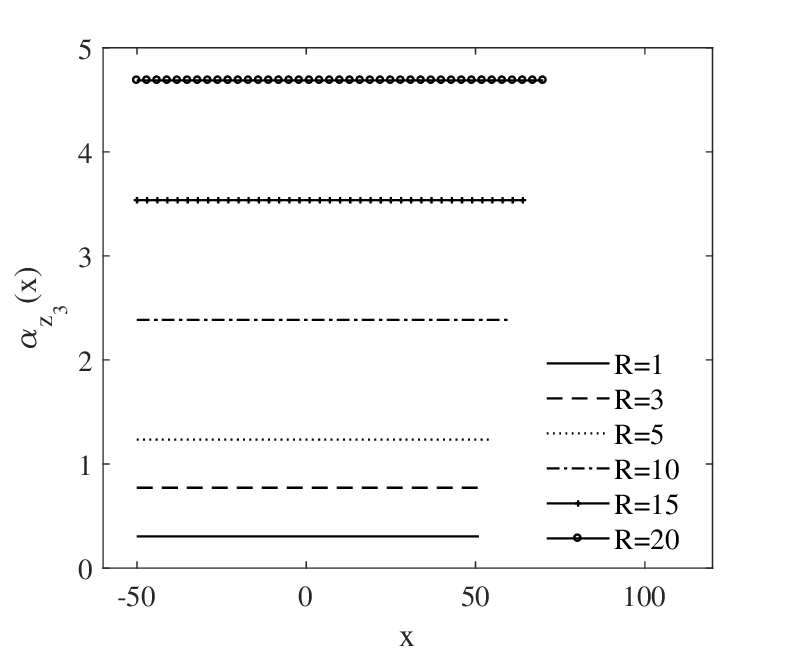}
\caption{The angle $\alpha(x)$ induced by different domains for the fundamental representation for various sizes of the Wilson loops in the Casimir scaling regime. The angle of the $z_0$ center vortices, $\alpha_{z_0}(x)$, changes very slowly as the sizes of the loops increase. The angle $\alpha_{z_1}(x)$ increases slowly while those for $z_2$ and $z_3$ center vortices change much more significantly. If the angle value grows very slowly with increasing sizes of the Wilson loops, deviations from the exact Casimir scaling would be less. This occurs because the potential ratios start out at the Casimir ratios. Therefore, the static potentials induced by $z_0$ center vortices are in the best agreement with the Casimir scaling. The free parameter is $L_{d}=100$ and $L^{2}_{d}/(2\mu)=4$.}\label{fig:11}
\end{figure}
In Fig. \ref{fig:11}, a comparison between the angles $\alpha(x)$ obtained from different domains for the fundamental representation is shown for various Wilson loop sizes in the Casimir scaling regime. The static potentials agree with the exact Casimir scaling for the loop with $R=1$ where the angle is small. As the loop sizes increase, the angle also increases. The angle corresponding to $z_0$ vacuum domains, $\alpha_{z_0}(x)$, increases very slowly in the Casimir scaling regime and therefore its potentials agree almost exactly with Casimir scaling. The angle corresponding to $z_1$ center vortices, $\alpha_{z_1}(x)$, increases slowly while those for $z_2$ and $z_3$ center vortices change rapidly in the Casimir scaling regime. Therefore, the deviations of the potentials induced by $z_2$ and $z_3$ center vortices from the exact Casimir scaling are much greater in the intermediate regime.    

Therefore, $z_2$ and $z_3$ center vortices could create concavity in the static potentials and break down the Casimir scaling in the intermediate regime. As explained by Faber $\it{et~ al.}$ \cite{Faber:1997rp}, it is possible that only vortices with the smallest magnitude of center flux have substantial probability. As shown in Fig. \ref{fig:4}, the group factors of $z_2$ and $z_3$ center vortices indicate that they are constructed of $z_1$ center vortices with the same flux orientations causing them to repel each other. These vortices are characterized by the center element $z_2=z_1^2$ and $z_3=z_1^3$. Therefore, $z_2$ and $z_3$ center vortices do not form stable configurations and each $z_1$ center vortex within them should be considered as a single vortex in the model. Consequently, the vacuum, instead of being filled with the $z_2$ and $z_3$ center vortices is filled with only $z_1$ vortices. As a result, repulsions in $z_4$ vacuum domain decompose it into $z_1$ center vortices. 

\begin{figure}[h!]
\centering
a)\includegraphics[width=0.42\columnwidth]{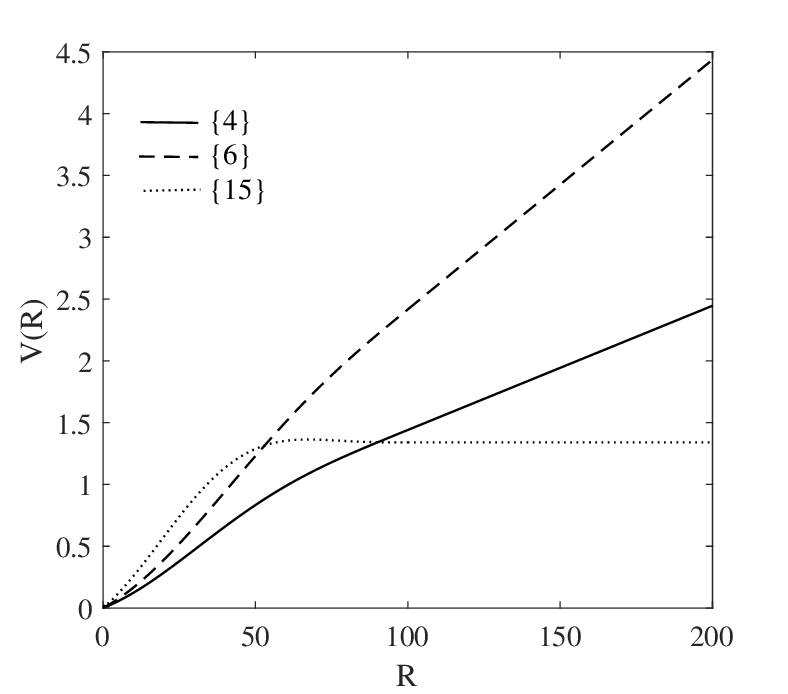}
b)\includegraphics[width=0.42\columnwidth]{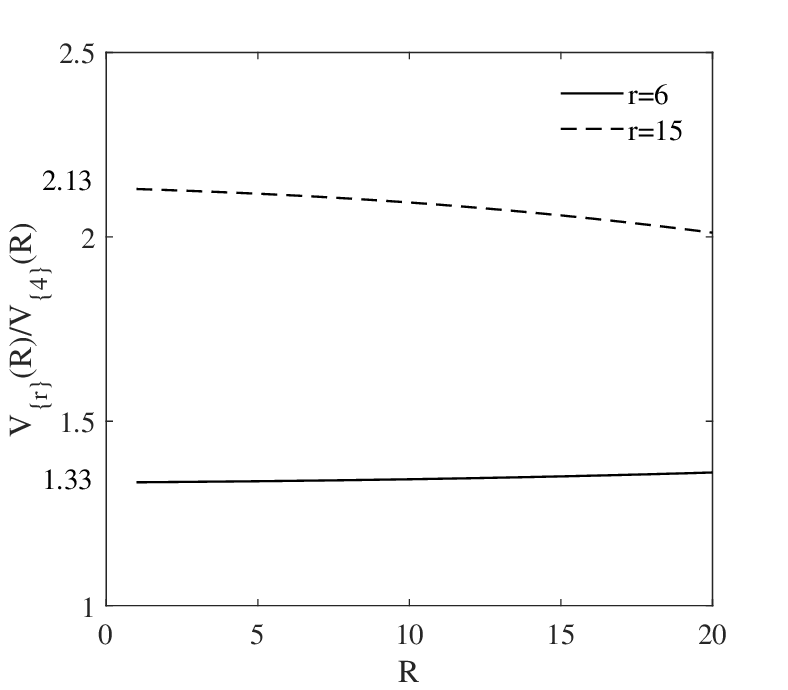}
\caption{a) The static potentials using $z_0$ and $z_1$ domains for the various representations of $SU(4)$. Repulsions within the $z_4$ vacuum domains can decompose them into $z_1$ center vortices with the lowest magnitude of center vortex fluxes. It seems that repulsions can lead to $N$-ality at large distances where there are three slopes of SU($4$) potentials. b) Potential ratios 
$V_{\{r\}}(R)/V_{\{4\}}(R)$ induced by $z_0$ and $z_1$ domains at the intermediate distances. Attractions within $z_0$ vacuum domains can improve Casimir scaling. The free parameters are $L_{d}=100$, $f_{0}=0.03$, $f_{1}=0.01$, and $L^{2}_{d}/(2\mu)=4$.}\label{fig:12}
\end{figure}

It seems that the center domains corresponding to $z_n$($n=1,2,3,4$) given in Eq. (\ref {center_elements}) deform to the lowest magnitude of center vortex fluxes, $z_1$ center vortices, due to repulsions. Additionally, attractions between the $z_1$ vortices and $z_1^*$ anti-vortices  lead to stable $z_0$ vacuum domains. As a result, in the Yang–Mills vacuum, there are $z_1$ center vortices as well as $z_0$ vacuum domains characterized by $z_1z_1^*$. Figure \ref{fig:12} a) plots the static potentials induced by $z_1$ center vortices and $z_0$ vacuum domains for the $\{4\}$, $\{6\}$ and $\{15\}$ representations for the range $R\in [0,200]$. 
As shown, in agreement with the tensor products of the Young diagrams given in Figs. \ref{fig:6} and \ref{fig:7}, there are three slops of the potentials at large distances where the one for the representation $\{15\}$ is zero value, indicating screening. Therefore, repulsions among $z_1$ vortices within $z_4$ vacuum domain lead to the agreement of the static potentials with $N$-ality at large distances.

Furthermore, the potential ratios $V_{\{6\}}(R)/V_{\{4\}}(R)$ and $V_{\{15\}}(R)/V_{\{4\}}(R)$ induced by $z_1$ center vortices and $z_0$ vacuum domains are shown in Fig. \ref{fig:12} b). Thus, the satisfactory static potentials for
the different representations can be achieved. The potential ratios start out at the Casimir ratios and drop very slowly from the exact Casimir scaling.

As a result, it seems that attractions between center vortices in $z_0$ vacuum domains lead to almost exact Casimir scaling in the static potentials at the intermediate regime while repulsions between center vortices in $z_4$ vacuum domains exhibit $N$-ality at the asymptotic regime.

\section{Conclusion}\label{Sect3}

We have studied the role of the vacuum domains corresponding to trivial center element in SU($4$) Yang-Mills theory within the framework of the domain structure model. We analyze two types of $z_0$ and $z_4$ vacuum domains to understand the basic properties of static potentials, Casimir scaling at the intermediate distances and $N$-ality at the large distances.
 
 The group factor $\mathcal{G}_r(\alpha^{(n)})$ for the two types of the vacuum domains shows that the $z_4$ vacuum domain is constructed of the SU($4$) center vortices with the same flux orientations and the $z_0$ vacuum domain can be formed by vortex and antivortex with opposite magnetic flux directions. Therefore, $z_4$ vacuum domain can be characterized by $z_1^4$, $z_2^2$, or $z_1z_3$ while the $z_0$ vacuum domain can be constructed with non trivial center elements as $z_nz_n^*$. 
 
Interactions within the vacuum domains may exhibit stable configurations. The QCD vacuum is a dual analogy to the type
II superconductivity where two vortices repel each other and the vortex-antivortex interaction is attractive. Therefore, vortices in the $z_4$ vacuum domain may repel each other while there is attraction between the vortex and anti-vortex in the $z_0$ vacuum domain.

 The potentials induced by $z_0$ vacuum domains satisfy almost the exact Casimir scaling at intermediate distances while there is color screening at large distances for all representations. It seems that using attractions between center vortices of $z_0$ vacuum domains, the static potentials strongly agree with the Casimir scaling at intermediate regimes but they destroy $N$-ality at large distances.

In addition, in the potentials induced by $z_4$ vacuum domains exhibit concavity, which is an artifact. The repulsions between center vortices in the $z_4$ vacuum domain suggest that the configuration is unstable. Therefore, each center vortex within the $z_4$ vacuum domain should be considered as a single vortex. Consequently, the vacuum in stead of the $z_4$ vacuum domains is filled with $z_1$, $z_2$, and $z_3$ vortices. Again, the concavity is appeared for the potentials induced by $z_2$ or $z_3$ center vortices and these configurations break down the Casimir scaling in the intermediate regime. The group factors of $z_2$ and $z_3$ center vortices indicate that they are characterized by the center element $z_2=z_1^2$ and $z_3=z_1^3$ and therefore it seems that these configurations are unstable and decompose into $z_1$ center vortices with the lowest magnitude of center vortex fluxes. 

As a result, SU($4$) Yang–Mills vacuum may contain only $z_1$ center vortices as well as $z_0$ vacuum domains characterized by $z_1z_1^*$. The static potentials induced by $z_1$ and $z_0$ center domains strongly satisfy Casimir scaling. Furthermore, we observe three slops of the potentials at large distances for the representations, in agreement with $N$-ality.

It seems that attractions between center vortices in $z_0$ vacuum domains lead to almost the exact Casimir scaling in the static potentials at the intermediate regime while repulsions between center vortices in $z_4$ vacuum domains exhibit $N$-ality at the asymptotic regime.

%


\begin{thebibliography}{99} 
\bibitem{Piccioni:2005un}
  C.~Piccioni,
  \textit{Casimir scaling in SU(2) lattice gauge theory},
  Phys.\ Rev.\  D {\bf 73}, 114509 (2006).
\bibitem{Deldar:1999vi} 
  S.~Deldar,
  \textit{Static SU(3) potentials for sources in various representations},
  Phys.\ Rev.\  D {\bf 62}, 034509 (2000).
\bibitem{Bali:2000un}
  G.~S.~Bali,
  \textit{Casimir scaling of SU(3) static potentials},
  Phys.\ Rev.\  D {\bf 62}, 114503 (2000).
\bibitem{Kratochvila:2003zj}
  S.~Kratochvila and P.~de Forcrand,
  \textit{Observing string breaking with Wilson loops},
  Nucl.\ Phys.\ B {\bf 671} (2003) 103.
\bibitem{Bachas:1985xs}
  C.~Bachas,
  \textit{Convexity of the Quarkonium Potential},
  Phys.\ Rev.\ D {\bf 33} (1986) 2723.  
\bibitem{DelDebbio:1996lih}
  L.~Del Debbio, M.~Faber, J.~Greensite and S.~Olejnik,
  \textit{Center dominance and Z(2) vortices in SU(2) lattice gauge theory},
  Phys.\ Rev.\ D {\bf 55} (1997) 2298.
\bibitem{Langfeld:1997jx}
  K.~Langfeld, H.~Reinhardt and O.~Tennert,
  \textit{Confinement and scaling of the vortex vacuum of SU(2) lattice gauge theory},
  Phys.\ Lett.\ B {\bf 419} (1998) 317
\bibitem{DelDebbio:1997ke}
  L.~Del Debbio, M.~Faber, J.~Greensite and S.~Olejnik,
  \textit{Center dominance, center vortices, and confinement},
  hep-lat/9708023.
\bibitem{Langfeld:1998cz}
  K.~Langfeld, O.~Tennert, M.~Engelhardt and H.~Reinhardt,
  \textit{Center vortices of Yang-Mills theory at finite temperatures},
  Phys.\ Lett.\ B {\bf 452} (1999) 301.
\bibitem{Engelhardt:1999fd}
  M.~Engelhardt, K.~Langfeld, H.~Reinhardt and O.~Tennert,
  \textit{Deconfinement in SU(2) Yang-Mills theory as a center vortex percolation transition},
  Phys.\ Rev.\ D {\bf 61} (2000) 054504.
\bibitem{Kovacs:1998xm}
  T.~G.~Kovacs and E.~T.~Tomboulis,
  \textit{Vortices and confinement at weak coupling},
  Phys.\ Rev.\ D {\bf 57} (1998) 4054.    
\bibitem{Dehghan:2024rly}
  Z.~Dehghan, R.~Golubich, R. H\"ollwieser, and M. Faber
  \textit{Investigation of the ensemble of maximal center gauge},
  Phys.\ Rev.\ D {\bf 110} (2024) 014501.   
\bibitem{Asmaee:2021xkm}
  Z.~Asmaee, S.~Deldar and M.~Kiamari,
  \textit{Introducing vortices in the continuum using direct and indirect methods},
  Phys.\ Rev.\ D {\bf 105} (2022) 096020.   
\bibitem{Golubich:2021kjc}
  R.~Golubich and M.~Faber,
  \textit{Properties of SU(2) Center Vortex Structure in Smooth Configurations},
  Particles {\bf 4} (2021) 93--105. 
\bibitem{Schweigler:2012ae}
 T.~Schweigler, R.~H\"ollwieser, M.~Faber, and Urs M.~Heller,
  \textit{Colorful SU(2) center vortices in the continuum and on the lattice},
  Phys.\ Rev.\ D {\bf 87} (2013) 054504.   
\bibitem{Hollwieser:2012kb}
 R.~H\"ollwieser, M.~Faber, and Urs M.~Heller,
  \textit{Critical analysis of topological charge determination in the background of center vortices in SU(2) lattice gauge theory},
  Phys.\ Rev.\ D {\bf 86} (2012) 014513.
\bibitem{tHooft:1977nqb}
  G.~'t Hooft,
  \textit{On the Phase Transition Towards Permanent Quark Confinement},
  Nucl.\ Phys.\ B {\bf 138} (1978) 1.
\bibitem{Vinciarelli:1978kp}
  P.~Vinciarelli,
  \textit{Fluxon Solutions in Nonabelian Gauge Models},
  Phys.\ Lett.\  {\bf 78B} (1978) 485.
\bibitem{Yoneya:1978dt}
  T.~Yoneya,
  \textit{$Z(N$) Topological Excitations in {Yang-Mills} Theories: Duality and Confinement},
  Nucl.\ Phys.\ B {\bf 144} (1978) 195.
\bibitem{Cornwall:1979hz}
  J.~M.~Cornwall,
  \textit{Quark Confinement and Vortices in Massive Gauge Invariant QCD},
  Nucl.\ Phys.\ B {\bf 157} (1979) 392.
\bibitem{Mack:1978rq}
  G.~Mack and V.~B.~Petkova,
  \textit{Comparison of Lattice Gauge Theories with Gauge Groups Z(2) and SU(2)},
  Annals Phys.\  {\bf 123} (1979) 442.
\bibitem{Nielsen:1979xu}
  H.~B.~Nielsen and P.~Olesen,
  \textit{A Quantum Liquid Model for the QCD Vacuum: Gauge and Rotational Invariance of Domained and Quantized Homogeneous Color Fields},
  Nucl.\ Phys.\ B {\bf 160} (1979) 380.      
\bibitem{Faber:1997rp}
  M.~Faber, J.~Greensite and S.~Olejnik,
  \textit{Casimir scaling from center vortices: Towards an understanding of the adjoint string tension},
  Phys.\ Rev.\ D {\bf 57} (1998) 2603.      
\bibitem{Deldar:2001}
  S.~Deldar,
  \textit{Potentials between static SU(3) sources in the fat-center-vortices
  model},
  \emph{JHEP} {\bf 0101} (2001) 013.  
\bibitem{Deldar:2007}  
  S.~Deldar and S.~Rafibakhsh,
  \textit{Confinement and the second vortex of the SU(4) gauge group},
  \emph{Phys.\ Rev.\/}  D {\bf 76} (2007) 094508.
\bibitem{Holland:2003jy}  
  K.~Holland, P.~Minkowski, M.~Pepe, and U.~J.~Wiese,
  \textit{Exceptional confinement in G$_2$ gauge theory},
  \emph{Nucl.\ Phys.\/}  B {\bf 668} (2003) 207.  
\bibitem{Greensite:2006sm}
  J.~Greensite, K.~Langfeld, S.~Olejnik, H.~Reinhardt and T.~Tok,
  \textit{Color Screening, Casimir Scaling, and Domain Structure in G(2) and SU(N) Gauge Theories},
  Phys.\ Rev.\ D {\bf 75} (2007) 034501.   
\bibitem{Olejnik:2009jr}  
  S.~Olejnik,
  \textit{Vacuum structure and Casimir scaling in Yang-Mills theories},
  \emph{PoS} {\bf QCD-TNT09} (2009) 032.    
\bibitem{Pepe:2007}  
  M.~Pepe and U.~J.~Wiese,
  \textit{Exceptional deconfinement in G$_2$ gauge theory},
  \emph{Nucl.\ Phys.\/}  B {\bf 768} (2007) 21.  
\bibitem{Liptak:2008gx}
  {L\kern-.08cm'}.~Lipt\'ak and {\v{S}}. Olejn{\'\i}k,
  \textit{Casimir scaling in G$_2$ lattice gauge theory},
  Phys.\ Rev.\  D {\bf 78} (2008) 074501.
\bibitem{Nejad:2014tja}
  S.~M.~Hosseini Nejad and S.~Deldar,
  \textit{Contributions of the center vortices and vacuum domain in potentials between static sources},
  JHEP {\bf 1503} (2015) 016.    
\bibitem{Nejad:2019}
  S.~M.~Hosseini Nejad,
  \textit{Domain structures and quark potentials in SU($3$) gauge theory},
  Phys.\ Rev.\  D {\bf 100} (2019) 114504.     
\bibitem{Engelhardt:1999wr}
  M.~Engelhardt and H.~Reinhardt,
  \textit{Center vortex model for the infrared sector of Yang-Mills theory: Confinement and deconfinement},
  Nucl.\ Phys.\ B {\bf 585} (2000) 591.
\bibitem{Engelhardt:2003wm}
  M.~Engelhardt, M.~Quandt and H.~Reinhardt,
  \textit{Center vortex model for the infrared sector of SU(3) Yang-Mills theory: Confinement and deconfinement},
  Nucl.\ Phys.\ B {\bf 685} (2004) 227.
\bibitem{Deldar:2010hw}
  S.~Deldar and S.~Rafibakhsh,
  \textit{Removing the concavity of the thick center vortex potentials by fluctuating the vortex profile},
  Phys.\ Rev.\ D {\bf 81} (2010) 054501.
\bibitem{Deldar:2011fh}
  S.~Deldar, H.~Lookzadeh and S.~M.~Hosseini Nejad,
  \textit{Confinement in G(2) Gauge Theories Using Thick Center Vortex Model and domain structures},
  Phys.\ Rev.\ D {\bf 85} (2012) 054501.
\bibitem{Deldar:2009aw}
  S.~Deldar and S.~Rafibakhsh,
  \textit{Short distance potential and the thick center vortex model},
  Phys.\ Rev.\ D {\bf 80} (2009) 054508.
\bibitem{Nejad:2014hka}
  S.~M.~Hosseini Nejad and S.~Deldar,
  \textit{Role of the SU($2$) and SU($3$) subgroups in observing confinement in the G($2$) gauge group},
  Phys.\ Rev.\ D {\bf 89} (2014) no.1,  014510.
\bibitem{Mandelstam} 
  S. Mandelstam,
  \textit{Vortices and quark confinement in non-abelian gauge theories},
  Phys. Reports 23C, 245–249 (1976).
\bibitem{Hooft}
G.~'t Hooft,
  \textit{High energy physics, in Gauge theories with unified weak,
electromagnetic, and strong interactions}, edited by A. Zichichi (EPS International Conference, Palermo, 1975).  
\bibitem{Chaves} 
  Andrey Chaves, F. M. Peeters, G. A. Farias, M. Milošević,
  \textit{Vortex-vortex interaction in bulk superconductors: Ginzburg-Landau theory},
  Phys.\ Rev.\ B {\bf 83}, 109905 (2011).
\bibitem{Kramer:1971zza} 
  L.~Kramer,
  \textit{Thermodynamic Behavior of Type-II Superconductors with Small kappa near the Lower Critical Field},
  Phys.\ Rev.\ B {\bf 3}, 3821 (1971).

\end{thebibliography}
\end{document}